\documentclass[prl,twocolumn,superscriptaddress,nofootinbib]{revtex4}
\usepackage[dvipdfmx]{graphicx}
\usepackage{color}
\usepackage{enumerate}
\usepackage{epsfig}
\usepackage{enumerate}
\usepackage{enumitem}
\usepackage{amsmath,amssymb,latexsym}
\usepackage{ascmac}
\usepackage{bm}
\newtheorem{theorem}{{\bf Theorem}}

\def\U#1{{\rm #1}}

\newcommand{\bra}[1]{\langle #1 |}
\newcommand{\ket}[1]{| #1 \rangle}

\newcommand{\expect}[1]{\left\langle #1 \right\rangle} 

\def\vac{\U{vac}}

\def\Pr{\U{Pr}}

\def\wt{\U{wt}}
\def\ph{\U{ph}}
\def\em{\U{em}}
\def\bit{\U{bit}}
\def\th{\U{th}}
\def\tr{\U{tr}}

\def\vac{{\rm vac}}

\begin{document}
\title{Quantum key distribution with any two independent and identically distributed states
}
\author{Akihiro Mizutani\\
{\it Mitsubishi Electric Corporation, Information Technology R\&D Center, 5-1-1 Ofuna, Kamakura-shi, Kanagawa, 247-8501 Japan}}

\begin{abstract}
To prove the security of quantum key distribution (QKD) protocols, several assumptions have to be imposed on users' devices. 
From an experimental point  of view, it is preferable that such theoretical requirements are feasible and the number of them is small. 
In this paper, we provide a security proof of a QKD protocol where the usage of {\it any} light source is allowed 
as long as it emits two independent and identically distributed (i.i.d.) states. 
Our QKD protocol is composed of two parts: the first part is characterization of the photon-number statistics of the emitted signals 
up to three-photons based on the method~[Opt. Exp. {\bf 27}, 5297 (2019)], followed by running our differential-phase-shift (DPS) protocol
~[npj Quantum Inf.~{\bf 5}, 87 (2019)]. 
It is remarkable that as long as the light source emits two i.i.d. states, 
even if we have no prior knowledge of the light source, we can securely employ it in the QKD protocol. 
As this result substantially simplifies the requirements on light sources, 
it constitutes a significant contribution on realizing truly secure quantum communication. 
\end{abstract}
\maketitle

\section{Introduction}
\label{sec_intro}
Quantum key distribution (QKD) enables two distant parties Alice and Bob to realize information-theoretically secure 
communication~\cite{LoNphoto2014}. 
To prove the security of the QKD protocol, we assume mathematical models on users' devices. 
It is preferable that such assumptions are experimentally feasible and the number of them is small; 
otherwise actual devices are hard to satisfy the theoretical requirements and the security of actual QKD systems cannot be guaranteed. 
The gap between the theoretical models of the actual devices and their physical properties is one of the major issues in the research 
field~\cite{Diamanti16}, and tremendous efforts have been made to relax the requirements on users' devices.
As for the measurement devices, we now have a practical and complete solution thanks to measurement-device-independent 
QKD~\cite{PhysRevLett.108.130503}. 
Therefore, the task left is securing the light sources. 

As for the light sources, 
since actual light sources never emit symmetrically-encoded single-photon states assumed 
in~\cite{PhysRevLett.85.441,Tomamichel2012,Tomamichel2017} nor 
perfectly phase-randomized coherent states supposed in~\cite{PhysRevLett.94.230504,PhysRevA.89.022307,Marcosfinite2014}, 
previous works have incorporated actual physical properties into the security proofs. 
For examples, QKD is shown to be secure with the relaxed assumptions such as basis independent states~\cite{PhysRevLett.90.057902}, 
discrete phase-randomized coherent states~\cite{1367-2630-17-5-053014}, non-phase-randomized coherent states~
\cite{lopreskill,PhysRevApplied.12.024048}, and asymmetrically-encoded states
~\cite{PhysRevA.74.042342,PhysRevA.90.052314,PhysRevA.92.032305,PhysRevLett.121.190502,AkihiroSCIC2018}. 
However, these previous works could not completely solve the aforementioned gap problem 
as experimentalists cannot verify whether these assumptions are really met in practice.

\begin{table}[h]
    \caption{Summary of the security proofs in which 	actual properties of the light sources are reflected. }
  \begin{tabular}{|c|c|c|} \hline
    Ref. & Protocol & Model of the light source\\ \hline \hline
    \cite{PhysRevLett.90.057902} & BB84 & Any basis-independent four states \\\hline
    \cite{lopreskill} & BB84 & Non-phase-randomized coherent states \\ \hline
    \cite{PhysRevApplied.12.024048} & 6-state & Non-phase-randomized coherent states \\\hline
    \cite{1367-2630-17-5-053014} & BB84 & Discrete phase-randomized coherent states \\\hline
    \cite{PhysRevA.74.042342,PhysRevA.90.052314,PhysRevA.92.032305,PhysRevLett.121.190502,AkihiroSCIC2018} &3-state &
Asymmetrically-encoded three states 
\\ \hline
    This work & DPS & Any two i.i.d states \\ \hline
  \end{tabular}
\label{tableI}
\end{table}
In this paper, we solve the gap problem for the light source that emits either state 
$\hat{\rho}_0$ or $\hat{\rho}_1$ depending on the input bit. 
See Table~\ref{tableI} for a summary of security proofs where the actual properties of the light sources are reflected. 
It is notable that our proof does not require any prior knowledge of the emitted signals. 
Although the security is guaranteed with any $\hat{\rho}_0$ and $\hat{\rho}_1$, it is obvious that the secret key rate depends on 
their states. 
It is easy to imagine that if $\hat{\rho}_0$ and $\hat{\rho}_1$ are orthogonal with each other, the security never be 
guaranteed. 
This implies that the expression of the key rate depends on some parameters that characterize $\hat{\rho}_0$ and $\hat{\rho}_1$. 
In our proof, we adopt the photon-number statistics of up to three photons for their characterizations, which 
can be obtained without any knowledge of $\hat{\rho}_0$ and $\hat{\rho}_1$ by using the 
method in~\cite{Kumazawa2017}. 
We note that our security framework covers a case where the light source is manufactured by a malicious party. 
Even under this situation, by checking the photon number statistics of the states emitted from the light source, 
we can securely employ it for QKD. 
After characterizing the photon-number statistics, Alice and Bob conduct our differential-phase-shift (DPS) QKD protocol~\cite{npjdps2019}, 
whose secret key rate depends on the characterized photon-number statistics. 
Note that, the only difference of our proof and the previous one~\cite{npjdps2019} lies in the assumption on Alice's light source, 
and Bob's measurement unit and the procedures of the protocol are exactly the same as those in~\cite{npjdps2019}. 
In proving the security of our protocol, we follow the same arguments done in~\cite{npjdps2019}. 

The paper is organized as follows. 
In section~\ref{sec_light}, we describe the model of the light source and explain the characterized parameters that come into the 
expression of the key rate. 
The setup and assumptions on Bob's measurement unit and the protocol description are exactly the same as those in our previous 
work~\cite{npjdps2019}. 
For self-consistency of this paper, we summarize them in sections~\ref{sec:modelmsn} and~\ref{sec_protocol}, respectively. 
In section~\ref{sec_proof}, we outline the security proof and state our main theorem, whose proof given in Appendix~\ref{appendixA}. 
In section~\ref{sec_simulation}, we present the simulation results of the key rate with coherent states. 
Finally, section~\ref{sec_conc} summarizes the paper.

\section{Model of light source}
\label{sec_light}
\begin{figure}[t]
\includegraphics[width=8cm]{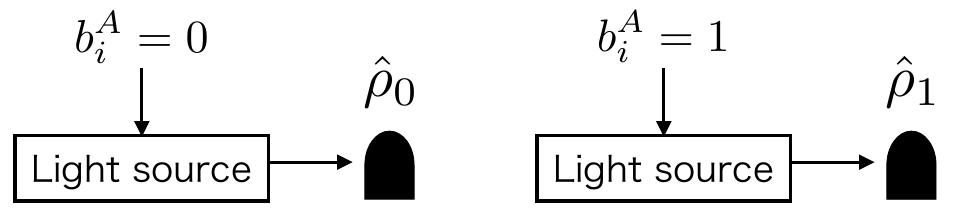}
\caption{
Our model of a light source with which we prove the security. 
The upper (lower) figure shows that when Alice chooses $b_i^A=0$ ($b_i^A=1$), the light source emits state $\hat{\rho}_0$ ($\hat{\rho}_1$).  
As long as the light source emits these i.i.d states, no other characterizations are required on the light source. 
}
\label{fig_iid}
        \end{figure}
\begin{figure}[t]
\includegraphics[width=8cm]{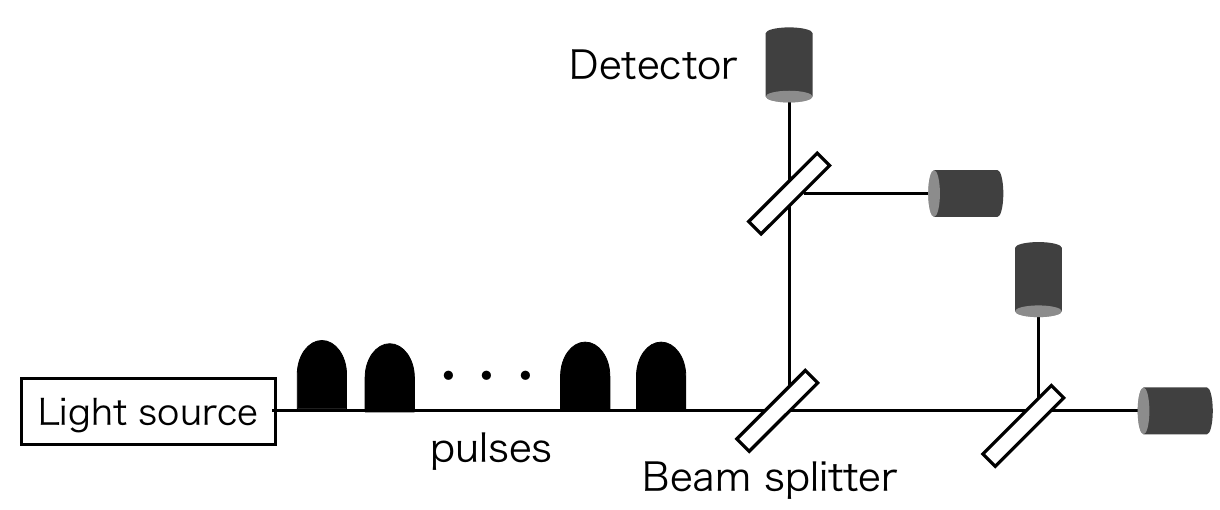}
\caption{
The characterization method~\cite{Kumazawa2017} using a Hanbury-Brown-Twiss setup with $D=4$ threshold detectors. 
By monitoring the $r$-fold coincidence probabilities for $r\in\{1,...,D\}$, Alice obtains all the parameters in $P_{\U{char}}$. 
Specifically, state $\hat{\rho}_0$ ($\hat{\rho}_1$) is measured for obtaining 
$p^L_0$ and $p^U_0$ ($p^L_1$ and $p^U_1$), and 
$\{\hat{\rho}^{\vec{b}_A}_S\}_{\vec{b}_A\in\{0,1\}^3}$ are measured for obtaining $\{q_n\}^3_{n=1}$. 
}
\label{fig_char}
        \end{figure}
In this section, we explain our model of the light source, which is illustrated in Fig.~\ref{fig_iid}.
Our model is based on the i.i.d. setting, where the light source emits state 
$\hat{\rho}_0$ ($\hat{\rho}_1$) when Alice chooses the bit 0~(1). 
In the case of our DPS QKD protocol~\cite{npjdps2019}, we consider that three optical pulses constitute a single-block from which 
we try to extract a one-bit key. 
The state of a single-block of three pulses is described below. 
\\\\
{\bf Description of the state of a single-block:}  
For each block, Alice chooses a random three-bit sequence $\vec{b}_A:=b^A_1b^A_2b^A_3\in\{0,1\}^3$, where $b^A_i$ is encoded only on 
  the $i^{\th}$ pulse in system $S_i$. 
  Depending on $\vec{b}_A$, Alice prepares a following state in system $S:=S_1S_2S_3$: 
  \begin{align}
    \hat{\rho}^{\vec{b}_A}_S:=\bigotimes^3_{i=1}\hat{\rho}^{b^A_i}_{S_i}.
\label{eq_model}
\end{align}
  Here, $\hat{\rho}^{b^A_i}_{S_i}$ denotes a density operator of the $i^{\th}$ pulse when $b^A_i$ is chosen. 
  We suppose that each system $R_i$ that purifies state $\hat{\rho}^{b^A_i}_{S_i}$ is possessed by Alice,
  and Eve does not have access to system $R_i$.
\\\\
We stress that no characterizations beyond the i.i.d. property are required on the light source. 
Obviously, the secret key rate depends on $\hat{\rho}_0$ and $\hat{\rho}_1$ (with $\hat{\rho}_{b^A_i}:=\hat{\rho}^{b^A_i}_{S_i}$), 
which can be understood by a following example. 
If the states emit more photons with higher probabilities, Eve obtains more information about the key because of the 
photon-number-splitting attack~\cite{PhysRevLett.85.1330}. 
This means that the key rate depends on some parameters characterizing $\hat{\rho}_0$ and $\hat{\rho}_1$. 
In our proof, we adopt the photon-number statistics of $\hat{\rho}_0$ and $\hat{\rho}_1$ 
up to three-photons as will be shown in Eqs.~(\ref{eq_pv0})-(\ref{eq_pron}). 
These characterized parameters are estimated before running the DPS protocol. 
\begin{enumerate}[label=(C\arabic*)]
\setlength{\parskip}{0cm}
\setlength{\itemsep}{0cm}
\item
{\bf Bounds on the vacuum emission probabilities:}
The vacuum emission probabilities of $\hat{\rho}_0$ and $\hat{\rho}_1$ are upper and lower bounded as
\begin{align}
p^L_0\le \tr\hat{\rho}_{0}\ket{\vac}\bra{\vac}\le p^U_0
\label{eq_pv0}
\end{align}
and
\begin{align}
p^L_1\le \tr\hat{\rho}_{1}\ket{\vac}\bra{\vac}\le p^U_1,
\label{eq_pv1}
\end{align}
respectively. Here, $\ket{\U{vac}}$ denotes the vacuum state. 
Note that the previous work~\cite{npjdps2019} assumes $\tr\hat{\rho}_{0}\ket{\vac}\bra{\vac}= \tr\hat{\rho}_1\ket{\vac}\bra{\vac}$. 
This is the exact difference in the assumption of the light source between this work and the previous one.
\item
{\bf Upper bounds on the tail distribution functions of the total photon number in a single-block:}
The probability that a single-block of pulses contains $n$ or more photons for $n\in\{1,2,3\}$ is upper-bounded. 
That is, for each $n$, 
\begin{align}
\sum^{\infty}_{m=n}\tr(\hat{\rho}_{b^A_1}\otimes\hat{\rho}_{b^A_2}\otimes\hat{\rho}_{b^A_3})\ket{m}\bra{m}\le q_n
\label{eq_pron}
\end{align}
holds for any $\vec{b}_A\in\{0,1\}^3$. 
Here, $\ket{m}$ is the $m$-photon number state in all the optical modes. 
\end{enumerate}
The characterized parameters 
\begin{align}
P_{\U{char}}:=\{p^L_0, p^U_0, p^L_1, p^U_1, q_1, q_2, q_3\}
\label{eq_char}
\end{align}
can be obtained by using the method~\cite{Kumazawa2017} that is based on the Hanbury-Brown-Twiss setup. 
This setup employs, for instance, 
$D=4$ threshold optical detectors and three beam splitters (whose setup is illustrated in Fig.~\ref{fig_char}), and 
by monitoring the $r$-fold coincidence probabilities for $r\in\{1,...,D\}$, 
tight bounds on each of all the parameters in $P_{\U{char}}$ can be obtained. 
The parameters $p^L_0$ and $p^U_0$ ($p^L_1$ and $p^U_1$) can be obtained by measuring $\hat{\rho}_{0}$ ($\hat{\rho}_1$), and 
$\{q_n\}^3_{n=1}$ are obtained by measuring $\{\hat{\rho}^{\vec{b}_A}_S\}_{\vec{b}_A\in\{0,1\}^3}$. 
Importantly, this method works for any i.i.d states.

After characterizing each of all the parameters in $P_{\U{char}}$, Alice and Bob conduct the DPS QKD protocol. 
The setup and assumptions on a measurement unit and the procedures of the QKD protocol are 
exactly the same as those in our previous work~\cite{npjdps2019}. 
For self-consistency of this paper, we summarize them in Secs.~\ref{sec:modelmsn} and \ref{sec_protocol}.

\section{Model of measurement unit}
\label{sec:modelmsn}
Here, we list up the assumptions on a measurement unit. 
\begin{enumerate}[label=(B\arabic*)]
\setlength{\parskip}{0cm}
\setlength{\itemsep}{0cm}
\item
Bob employs a one-bit delay Mach-Zehnder interferometer with two 
50:50 beam splitters (BSs), whose delay is equal to the interval of the neighboring emitted pulses.
\item
  After pulses pass through the second BS, the pulses are detected by two 
  photon-number-resolving (PNR) detectors that discriminate the vacuum, a single-photon, and two or more photons 
of a specific optical mode
\footnote{
One can extend our proof introduced in this paper to the use of threshold detectors by following the argument in~\cite{Sasaki2017QST}. 
It is based on the idea of monitoring the number of double-click events occurring in a single-block of pulses. 
But to estimate this quantity, we need to put an optical shutter in a long arm of the Mach-Zehnder interferometer.
}. 
Bob obtains the bit 0 or 1 according to which detector has reported the detection. 
  We assume that the quantum efficiencies and dark countings are the same for both detectors. 
\end{enumerate}
Bob detects photons at time slot 0 through 3, where the $j^{\th}$ ($1\le j\le 2$) time slot is defined as an 
expected detection time from the superposition of the $j^{\th}$ and $(j+1)^{\th}$ incoming pulses, 
and the $0^{\th}$ ($3^{\U{rd}}$) time slot is defined as an expected detection time 
from the superposition of the 1$^{\U{st}}$ ($3^{\U{rd}}$) incoming pulse and the $3^{\U{rd}}$ incoming pulse 
in the previous block (1$^{\U{st}}$ incoming pulse in the next block)
\footnote{
Note that we describe the protocol where the emitted pulses are separated at the equal time interval. 
However, our security analysis, which will be given in Sec.~\ref{sec_proof}, can be applied even if Alice emits pulses with different time intervals. 
This is because the security analysis is focused only on each single-block of pulses, 
and no key-bit is extracted from the interfered pulses coming from different blocks.
}
.

\section{Protocol description}
\label{sec_protocol}
We describe the procedures of the protocol. In its description, $|\bm{\kappa}|$ denotes the length of a bit sequence $\bm{\kappa}$.
\begin{enumerate}[label=(P\arabic*)]
\setlength{\parskip}{0cm}
\setlength{\itemsep}{0cm}
\item
  Alice chooses a random three-bit sequence $\vec{b}_A\in\{0,1\}^3$ and sends the corresponding three-pulse state 
$\hat{\rho}^{\vec{b}_A}_S$ to Bob through a quantum channel.
  \item
	We define the event {\it detected} if the PNR detectors have reported one photon in total among the $1^{\U{st}}$ and $2^{\U{nd}}$ time slots. 
	Depending on which detector has reported the detection at the $j^{\th}$ ($1\le j\le 2$) time slot, Bob obtains the raw key bit  $k_B\in\{0,1\}$. 
	If the detected event does not occur, Alice and Bob skip steps (P3) and (P4). 
  \item
    Bob announces $j$ via an authenticated public channel.
  \item
    Alice calculates her raw key bit $k_A=b^A_j\oplus b^A_{j+1}\in\{0,1\}$.
  \item
    Alice and Bob repeat (P1)-(P4) for $N_{\U{em}}$ times. 
  \item
    Alice randomly selects a small portion of her raw key as sampled bits for estimating the bit error rate $e_{\bit}$ among them. 
	Through the authenticated public channel, Alice and Bob compare the bit values of the sampled bits and obtain $e_{\bit}$. 
	Using $e_{\bit}$, the bit error rate in the remaining raw key can be obtained. 
\item 
By concatenating their remaining raw keys, Alice and Bob obtain their sifted keys $\bm{\kappa}_A$ and $\bm{\kappa}_B$, respectively. 
  \item
  Bob performs bit error correction on his sifted key $\bm{\kappa}_B$ by sacrificing the bits $|\bm{\kappa}_A|f_{\U{EC}}$ 
of encrypted public communication from Alice by consuming the same length of a pre-shared secret key.
  \item
    Alice and Bob perform privacy amplification by shortening the fraction $f_{\U{PA}}$ to obtain the final keys.
\end{enumerate}
 In this paper, we only consider the asymptotic secret key rate, where 
 the following observed parameters are fixed in the limit of $N_{\em}\to\infty$. 
\begin{align}
  0\le Q:=\frac{|\bm{\kappa}_A|}{N_{\U{em}}}\le1,~~0\le e_{\U{bit}}\le 1.
  \label{observed}
  \end{align}

\section{Security proof}
\label{sec_proof}
In this section, we summarize the security proof of our protocol and determine the amount of privacy amplification $Qf_{\U{PA}}$. 
\subsection{Notations}
In what follows, we adopt the following notations. 
The function $h(x)$ is defined as 
\begin{align}
h(x)=  
\begin{cases}
    -x\log_2x-(1-x)\log_2(1-x)~~ (0\le x\le 0.5)\\
    1~~(x>0.5).
  \end{cases}
\end{align}
The Hadamard operator $\hat{H}$ is denoted by 
\begin{align}
\hat{H}=1/\sqrt{2}\sum_{x,y=0,1}(-1)^{xy}\ket{x}\bra{y}.
\end{align}
The $Z$-basis and $X$-basis states for $j^{\U{th}}$ qubit system $A_j$ are defined as $\{\ket{0}_{A_j},\ket{1}_{A_j}\}$ 
and $\{\ket{+}_{A_j},\ket{-}_{A_j}\}$ with $\ket{\pm}_{A_j}=(\ket{0}_{A_j}\pm\ket{1}_{A_j})/\sqrt{2}$, respectively. 
The controlled-not (CNOT) gate $\hat{U}^{(j)}_{\U{CNOT}}$ on the $Z$-basis is defined as 
\begin{align}
\hat{U}^{(j)}_{\U{CNOT}}\ket{x}_{A_j}\ket{y}_{A_{j+1}}=\ket{x}_{A_j}\ket{x+y~\U{mod}2}_{A_{j+1}}
\end{align}
with $x,y\in\{0,1\}$.

\subsection{Summary of security proof}
Here, we summarize our security proof and state the main theorem that derives the 
amount of privacy amplification $Qf_{\U{PA}}$. 
The proof of Theorem~\ref{mainth} is given in Appendix~\ref{appendixA}. 
For our security proof, we follow the same arguments done in our previous work~\cite{npjdps2019}, 
which employs complementarity~\cite{Koashi2009}. 
The security proof starts from substituting the following actual steps with the virtual ones. 
\begin{enumerate}[label=$\bullet$]
\setlength{\parskip}{0cm}
\setlength{\itemsep}{0cm}
\item
Alice's state preparation in step~(P1) 
\item
Alice's calculation of her raw key bit $k_A$ in step~(P4)
\end{enumerate}
Regarding Alice's state preparation in step~(P1),
she alternatively prepares three auxiliary qubit systems $A_1$, $A_2$ and $A_3$ that remain at Alice's site and are entangled with 
actual states in system $S$. Specifically, Alice virtually prepares the following state
\begin{align}
  \ket{\Phi}_{ASR}:=
  2^{-3/2}\bigotimes^3_{i=1}\sum^1_{b^A_i=0}\hat{H}\ket{b^A_i}_{A_i}\ket{\psi_{b^A_i}}_{S_iR_i}.
  \label{coherentLstates}
\end{align}
Here, $\ket{\psi_{b^A_i}}_{S_iR_i}$ is a purification of $\hat{\rho}^{b^{A}_i}_{S_i}$, that is,
$\tr_{R_i}\ket{\psi_{b^A_i}}\bra{\psi_{b^A_i}}_{S_iR_i}=\hat{\rho}^{b^{A}_i}_{S_i}$.
Recall that system $R_i$ is assumed to be possessed by Alice. 

As for calculation of Alice's raw key bit $k_A$ in step~(P4), she alternatively performs the CNOT gate $\hat{U}^{(j)}_{\U{CNOT}}$ 
on the qubits in systems $A_j$ and $A_{j+1}$ with the $j^{\th}$ qubit being the control and the $(j+1)^{\th}$ one being the target. 
After performing the CNOT gate, she measures the $j^{\th}$ auxiliary qubit in the $X$-basis to obtain $k_A$. 

For evaluating the security of the key $\bm{\kappa}_A$, we consider the complementary scenario~\cite{Koashi2009}. 
We define the complementary observable to the one to obtain $k_A$ as the $Z$-basis measurement on the $j^{\U{th}}$ auxiliary 
qubit after performing $\hat{U}^{(j)}_{\U{CNOT}}$. 
Let $z_j$ denote the outcome of the $Z$-basis measurement on the  $j^{\U{th}}$ qubit, and we quantify how well Alice can predict $z_j$. 
To enhance the accuracy of her prediction of $z_j$, she employs the following information. 
\begin{enumerate}[label=$\bullet$]
\setlength{\parskip}{0cm}
\setlength{\itemsep}{0cm}
\item
The $Z$-basis measurement outcome on the $(j+1)^{\U{th}}$ qubit after performing $\hat{U}^{(j)}_{\U{CNOT}}$.  
\item
Bob's virtual measurement to learn which of the $j^{\U{th}}$ or $(j+1)^{\U{th}}$ time slot has a single-photon. 
\end{enumerate}
We define the occurrence of a phase error to be the case where Alice fails her prediction.  
The explicit formula of the POVM element of obtaining a phase error is expressed in Eq.~(\ref{phPOVM}) in Appendix~\ref{appendixA}. 
When $N_{\ph}$ denotes the number of phase errors, that is, the number of wrong predictions on the outcome $z_j$ among $|\bm{\kappa}_A|$ trials, 
the phase error rate is expressed as $e_{\ph}=N_{\ph}/|\bm{\kappa}_A|$. 
Once we obtain the upper bound $f(Q, e_{\bit}, P_{\U{char}})$ on $N_{\ph}$ as a function of  experimentally observed parameters 
$Q$, $e_{\bit}$ in Eq.~(\ref{observed}) and $P_{\U{char}}$ in Eq.~(\ref{eq_char}), a sufficient amount of privacy amplification 
in the asymptotic limit of large key length $|\bm{\kappa}_A|$ is given by~\cite{Koashi2009}
\begin{align}
Qf_{\U{PA}}=Qh\left(\frac{f(Q, e_{\bit}, P_{\U{char}})}{|\bm{\kappa}_A|}\right).
\end{align}
Then, the secret key rate per a block is given by
\begin{align}
  R=Q\left[1-f_{\U{EC}}-h\left(\frac{f(Q, e_{\bit}, P_{\U{char}})}{|\bm{\kappa}_A|}\right)\right]/3.
  \label{keyrate}
\end{align}
The quantity $e^{\U{U}}_{\U{ph}}:=f(Q, e_{\bit}, P_{\U{char}})/|\bm{\kappa}_A|$ in Eq.~(\ref{keyrate}) is the upper bound on 
the phase error rate $e_{\U{ph}}$.
Our main result, Theorem~\ref{mainth}, derives $e^{\U{U}}_{\ph}$ with experimentally available parameters $Q$, $e_{\bit}$ 
and $P_{\U{char}}$. 
\begin{theorem}
In  the asymptotic limit of large key length $|\bm{\kappa}_A|$, the upper bound on the phase error rate of the DPS protocol is given by
\begin{align}
    e^{\U{U}}_{\ph}=(3+\sqrt{5})e_{\bit}+\frac{(3+\sqrt{5})\sqrt{s^U_1s^U_3}+s^U_{\ge2}}{Q},
\label{maineq}
\end{align}
where 
\begin{align}
s^U_3&:=q_3+t^3+6t^2+3t,
\label{eq_sU3}\\
s^U_1&:=q_1+3t,
\label{eq_sU1}\\
s^U_{\ge2}&:=q_2+t^3+9t^2+6t
\label{eq_sU2}
\end{align}
with
\begin{align}
t:=\frac{\max\left\{(\sqrt{p^U_0}-\sqrt{p^L_1})^2,(\sqrt{p^L_0}-\sqrt{p^U_1})^2\right\}}{4}.
\label{deft}
\end{align}
\label{mainth}
\end{theorem}
Note that the previous work~\cite{npjdps2019} corresponds to the case of $t=0$ as the vacuum emission probabilities are assumed to 
be equivalent for both bits. 
By substituting $t=0$ to Eq.~(\ref{maineq}), we recover the expression of the phase error rate in~\cite{npjdps2019}. 

\section{Simulation of secret key rate}
\label{sec_simulation}
\begin{figure}[t]
\includegraphics[width=8.5cm]{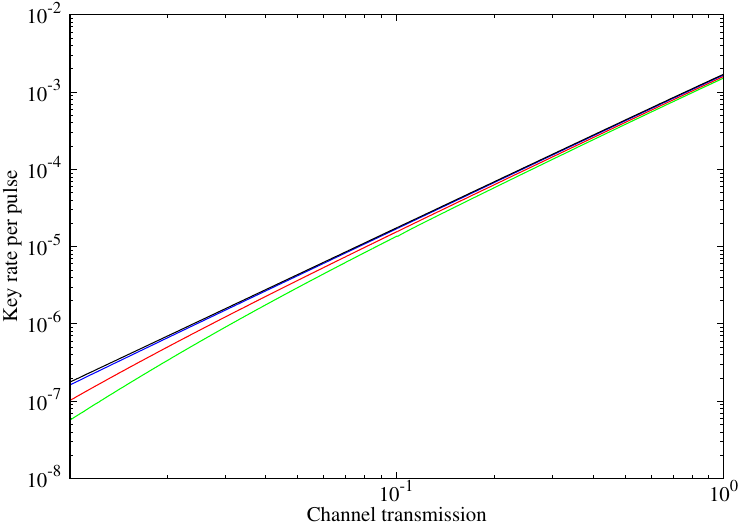}
\caption{
Secret key rate $R$ per pulse as a function of the overall channel transmission $\eta$. 
From top to bottom, we plot the key rates for the cases of 0, 1, 3, and 5\% intensity fluctuations 
under $e_{\bit}=1\%$. 
Note that the top line with no-intensity fluctuations corresponds to the case of~\cite{npjdps2019}. 
}
\label{fig_keyrate}
        \end{figure}
\begin{figure}[t]
\includegraphics[width=8.5cm]{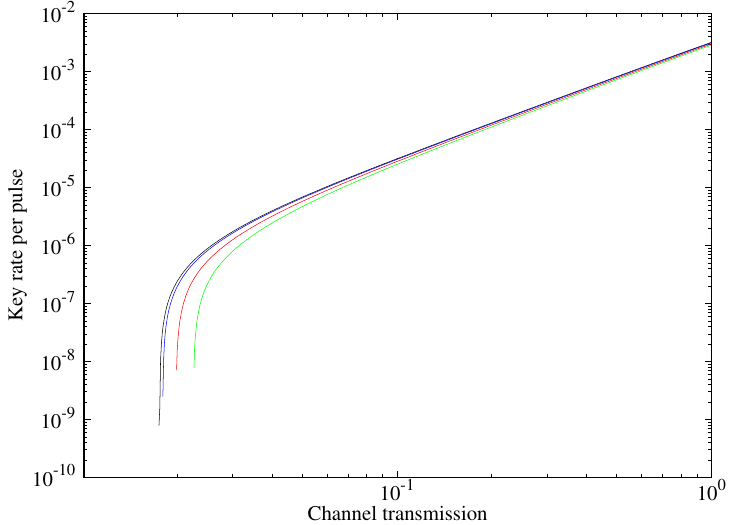}
\caption{
Secret key rate $R$ per pulse as a  function of the overall channel transmission $\eta$. 
From top to bottom, we plot the key rates for the cases of 0, 1, 3 and 5$\%$ intensity fluctuations.
Unlike the calculations in Fig.~\ref{fig_keyrate}, $e_{\U{bit}}$ is not constant, and the degradation of $e_{\U{bit}}$ over distance 
is reflected.
}
\label{fig_keyrate2}
\end{figure}

In this section, we show the simulation results of the key rate as a function of the 
channel transmittance $\eta$ including the detection efficiency. 
For the simulation purpose, we suppose that each emitted pulse is a coherent pulse; 
the states $\hat{\rho}_0$ and $\hat{\rho}_1$ are assumed to be given by 
$\hat{\rho}_0=\ket{\sqrt{\mu_0}}\bra{\sqrt{\mu_0}}$ and $\hat{\rho}_1=\ket{\sqrt{\mu_1}}\bra{\sqrt{\mu_1}}$, respectively.  
Here, $\ket{\sqrt{\mu}}:=e^{-\frac{\mu}{2}}\sum^{\infty}_{n=0}\frac{\mu^{\frac{n}{2}}}{\sqrt{n!}}\ket{n}$. 
We consider that the mean photon numbers $\mu_0$ and $\mu_1$ fluctuate by $a\%$ 
from an expected one $\mu$, namely, $\mu_0$ and $\mu_1$ lie in the range $[(1-0.01a)\mu, (1+0.01a)\mu]$. 
Note that we do not assume that $\mu_0=\mu_1$. 
In this case, the bounds on the vacuum emission probabilities are written as 
\begin{align}
p^U_0=p^U_1=e^{-(1-0.01a)\mu}
\end{align}
and
\begin{align}
p^L_0=p^L_1=e^{-(1+0.01a)\mu},
\end{align}
which results in 
\begin{align}
t=\frac{\left(\sqrt{e^{-(1-0.01a)\mu}}-\sqrt{e^{-(1+0.01a)\mu}}\right)^2}{4}.
\end{align}
The upper bound on the probability that a single block of three pulses contains $n$ or more photons for $n\in\{1,2,3\}$ is written as
\begin{align}
q_n=\sum^{\infty}_{\nu=n}e^{-3\mu(1+0.01a)}[3\mu(1+0.01a)]^{\nu}/\nu!.
\end{align}
In the simulation, we suppose $f_{\U{EC}}=h(e_{\bit})$ and the detection rate $Q=2\eta\mu e^{-2\eta\mu}$. 
With this setup, in Fig.~\ref{fig_keyrate}, we show the results of the key rate by setting $e_{\bit}=1\%$ and by varying the 
amount of intensity fluctuations of $a=0, 1, 3$ and 5. 
Only the top line with $a=0$ can be obtained by the previous work~\cite{npjdps2019}. In the other cases ($a$=1, 3 and 5), 
the vacuum emission probabilities are different for each bit, and $t$ is non-zero. 
As $t$ is a monotonically increasing function of $a$, the key rates in Fig.~\ref{fig_keyrate} degrade as the intensity fluctuations increase. 
Fig.~\ref{fig_keyrate} reveals that even if the vacuum emission probabilities fluctuate for both bits, 
such fluctuations do not have a significant impact on the degradation of the key rate. 

In the simulation of the key rate shown in Fig.~\ref{fig_keyrate}, the bit error rate $e_{\U{bit}}$ is set to be constant against channel transmission.
In practice, as this is not the case,  we next consider the $\eta$-dependent bit error rate by assuming the non-zero dark count error rate 
$d=10^{-7}$ of the detectors.
If we set the detection rate $Q$ as the probability of obtaining a successful detection among the 1$^{\U{st}}$ and 2$^{\U{nd}}$ time slots, 
it is given by
\begin{align} 
Q=2\left[e^{-\mu\eta}\mu\eta(1-d)+2de^{-\mu\eta}(1-d)\right]\left[e^{-\mu\eta}(1-d)^2\right].
\label{modQ}
\end{align}
Here, the first braket is the probability of obtaining a successful detection in the 1$^{\U{st}}$ (2$^{\U{nd}}$) time slot, 
and the second one is the probability of obtaining a no-click event in the 2$^{\U{nd}}$ (1$^{\U{st}}$) time slot.
Note that by substituting $d=0$, Eq.~(\ref{modQ}) derives the detection rate in Fig.~\ref{fig_keyrate}. 
Also, we suppose that $e_{\U{bit}}$ as the probability of obtaining a successful detection resulting in the bit error conditioned 
on the successful detection. It is written as 
\begin{align} 
e_{\U{bit}}=2\left[e^{-\mu\eta}d(1-d)\right]\left[e^{-\mu\eta}(1-d)^2\right]/Q.
\end{align}
Here, the first braket is the probability of obtaining a successful detection in a detector resulting in the bit error in the 1$^{\U{st}}$ 
(2$^{\U{n d}}$) time slot, and the second one is the probability of obtaining a no-click event in the 2$^{\U{nd}}$ (1$^{\U{st}}$) time slot.
Except for $d>0$, any other setups are the same as Fig.~\ref{fig_keyrate}.
In Fig.~\ref{fig_keyrate2}, we show the simulation result of the key rate with intensity fluctuations of $a = 0,1,3$ and 5. 
From Fig.~\ref{fig_keyrate2}, we find a similar tendency as shown in Fig.~\ref{fig_keyrate}, where the fluctuation of the intensity 
does not significantly degrade the key rate.

\section{Conclusion and discussion}
\label{sec_conc}
In this paper, we have provided an information-theoretic security proof of the DPS QKD protocol with {\it any} two i.i.d. states. 
More specifically, we have relaxed the assumptions on Alice's light source in~\cite{npjdps2019} to take into account a practical imperfection 
where the vacuum emission probabilities are different for each bit. 
The feature of our QKD protocol is that it starts from characterizing the photon-number statistics of the emitted signals up to three-photons 
based on the characterization method~\cite{Kumazawa2017}. 
After the photon-number statistics are estimated, 
our DPS protocol~\cite{npjdps2019} is conducted, whose secret key rate depends on the characterized statistics. 
Since this work significantly expands the light sources that can be securely employed in the QKD protocol, 
it paves a significant step toward truly secure quantum communication. 

We here summarize the differences in the security and performance between this work and the previous one~\cite{npjdps2019}. 
If the vacuum emission probabilities are different for each bit, the bounds on these probabilities in Eqs.~(\ref{eq_pv0}) and (\ref{eq_pv1}) 
come into the expression of the key rate through the parameter $t$ in Eq.~(\ref{deft}). 
If the estimation of these bounds are loose, we find from Eqs.~(\ref{maineq}) and (\ref{deft}) that the resulting phase error rate increases. 
This is the reason why the key rates in Figs.~\ref{fig_keyrate} and \ref{fig_keyrate2} degrade as the intensity fluctuations of the light sources 
increase. 
Note that only the top lines in Figs.~\ref{fig_keyrate} and \ref{fig_keyrate2} can be obtained using the result of~\cite{npjdps2019}
as the vacuum emission probabilities are the same for both bits. But the rest of the lines cannot be obtained with the previous result.

We end this section with two future works. 
The identicalness assumption supposed in our model of the light source is not necessary. 
Even if the light source emits non-identical states, as long as Eqs.~(\ref{eq_pv0}), (\ref{eq_pv1}) and (\ref{eq_pron}) hold for any state, 
the same argument holds. 
The independence assumption, on the other hand, is a crucial one for proving the security. 
It might be possible to remove this assumption by using our recent technique in~\cite{margarida2019} that can incorporate 
correlations among the emitted pulses. However, we leave this question in a future work. 

We next explain the other future work. In this paper, we regard consecutive three emitted pulses as a block. 
This means that the $(3n-2)^{\U{th}}$, $(3n-1)^{\U{th}}$ and $3n^{\U{th}}$ $(n\in\mathbb{N})$ 
emitted pulses are proactively assigned to the $n^{\U{th}}$ block. 
However, it is possible to retroactively assign the grouping of three pulses after 
Bob's detections to maximize the key rate as long as assumption~(C2) holds for any assigned three pulses. 
We leave the question in a future work about how much improvement could be obtained by this modification.

\section*{Acknowledgments}
We thank Prof. Kiyoshi Tamaki for helpful discussions.

\appendix
\section{Proof of Theorem~\ref{mainth}}
\label{appendixA}
\subsection{Notations}
  We first summarize the notations used in this Appendix. 
  \begin{align}
\hat{P}[\ket{\psi}]:=\ket{\psi}\bra{\psi}
  \end{align}
  for a vector $\ket{\psi}$ that is not necessarily normalized, and the Kronecker delta
    \begin{align}
\delta_{x,y}:=
\begin{cases}
  1 & x=y\\
  0 & x\neq y.
  \end{cases}
    \end{align}
The $Z$-basis states of Alice's auxiliary qubit system $A:=A_1A_2A_3$ is defined as
\begin{align}
  \ket{\bm{z}}_A:=\bigotimes^3_{i=1}\ket{z_i}_{A_i}
  \end{align}
with $\bm{z}:=z_1z_2z_3$ and $z_i\in\{0,1\}$, and $\wt(\bm{z})$ denotes the Hamming weight of a bit string $\bm{z}$. 
Furthermore, we define the projector $\hat{P}_a$ (with $0\le a \le 3$) as
\begin{align}
\hat{P}_a:=\sum_{\bm{z}:\wt(\bm{z})=a}\hat{P}[\ket{\bm{z}}_A].
\end{align}

\subsection{POVM elements}
Here, we introduce the formulas of POVM elements for detecting the bit $k_B$ at the $j^{\U{th}}$ time slot when the detected event occurs, 
the occurrence of bit error and phase error. 
Each of all the formulas are exactly the same as those in~\cite{npjdps2019}, and all the detailed derivations are referred to~\cite{npjdps2019}. 

We first introduce the POVM elements that correspond to detecting the bit $k_B$ at the $j^{\U{th}}$ ($j\in\{1,2\}$) 
time slot when the detected event occurs at step~(P2) in the actual protocol. 
For this, we consider the following procedures to learn whether the event is detected or not prior to determining $k_B$ and $j$. 
Bob puts the first (third) pulse to the first beam splitter in Bob's measurement unit and keeps the pulse passing through the long (short) arm, 
which we call first half pulse (third half pulse), while he keeps the second pulse as it is. 
Then, Bob performs the quantum non-demolition (QND) measurement to learn the total photon number among the first and 
third half pulses and the second pulse. 
The detected event is equivalent to obtaining one-photon in this measurement. 
If the QND measurement reveals one-photon, the state of Bob's system is spanned by the orthonormal basis 
$\{\ket{i}_B\}^3_{i=1}$ with $\ket{i}_B$ 
denoting the position of the single-photon at the first (third) half pulse for $i=1~(3)$ and at the original second pulse for $i=2$. 
If the detected event occurs, 
the POVM element $\{\hat{\Pi}_{j,k_B}\}_{j,k_B}$ for detecting the bit $k_B$ at the $j^{\th}$ time slot ($1\le j\le 2$) is given by
  \begin{align}
    \hat{\Pi}_{j,k_B}=\hat{P}[\ket{\Pi_{j,k_B}}_B]
  \end{align}
  with
  \begin{align}
\ket{\Pi_{j,k_B}}_B&:=\frac{\sqrt{w_j}\ket{j}_B+(-1)^{k_B}\sqrt{w_{j+1}}\ket{j+1}_B}{\sqrt{2}}.
\end{align}
Here, $w_1=w_3=1$ and $w_2=1/2$. 

Next, we show the formulas of the bit and phase error POVM elements $\hat{e}_{\bit}$ and $\hat{e}_{\ph}$. 
Importantly, these are defined only on systems $A$ and $B$, and the assumptions on 
emitted system $S$ does not come into their expressions. 
Since the only difference between our proof and the one in~\cite{npjdps2019} lies in 
whether $\tr\hat{\rho}_0\ket{\U{vac}}\bra{\U{vac}}=\tr\hat{\rho}_1\ket{\U{vac}}\bra{\U{vac}}$ is assumed or not, 
we can employ the same expressions of $\hat{e}_{\bit}$ and $\hat{e}_{\ph}$ as those in~\cite{npjdps2019}. 
From~\cite{npjdps2019}, we have 
  \begin{align}
    \hat{e}_{\U{bit}}=\sum^{2}_{j=1}\hat{e}^j_{\U{bit}},~\hat{e}_{\U{ph}}=\sum^{2}_{j=1}\hat{e}^j_{\U{ph}}
    \label{ebitephOperator}
  \end{align}
with
\begin{align}
  &\hat{e}^{j}_{\U{bit}}=(\hat{P}[\ket{++}_{A_jA_{j+1}}]+\hat{P}[\ket{--}_{A_jA_{j+1}}])\otimes\hat{\Pi}_{j,1}\notag\\
  &+(\hat{P}[\ket{+-}_{A_jA_{j+1}}]+\hat{P}[\ket{-+}_{A_jA_{j+1}}])\otimes\hat{\Pi}_{j,0}
  \label{bitpovm}
  \end{align}
and
  \begin{align}
    &\hat{e}^j_{\U{ph}}=\sum_{\bm{z}}\hat{P}[\ket{\bm{z}}_A]\notag\\
    \otimes&\left[w_j\delta_{z_{j+1},1}\hat{P}[\ket{j}_B]+
      w_{j+1}\delta_{z_j,1}\hat{P}[\ket{j+1}_B]\right].
    \label{phPOVM}
  \end{align}
When $\hat{\sigma}$ denotes a state of Alice and Bob's systems $A$ and $B$ just after the QND measurement reveals 
exactly one photon, the probability of having a bit (phase) error is given by $\tr\hat{\sigma}\hat{e}_{\bit}$ 
($\tr\hat{\sigma}\hat{e}_{\ph}$).

\subsection{Relation between bit and phase error rates}
Here, we derive the relation between the bit error rate $e_{\bit}$ and the phase error rate $e_{\ph}$. 
For this, we first employ Lemmas~1 and 2 in~\cite{npjdps2019} and obtain the following 
relation between the probabilities of having a phase error and a bit error, which holds for any $\hat{\sigma}$: 
\begin{align}
\tr\hat{e}_{\ph}\hat{\sigma}\le \lambda\left(
\tr\hat{e}_{\bit}\hat{\sigma}+\sqrt{(\tr\hat{\sigma}\hat{P}_1)(\tr\hat{\sigma}\hat{P}_3})\right)+\sum^3_{a=2}\tr\hat{P}_a\hat{\sigma}.
\label{main_pro}
\end{align}
Here, $\lambda:=3+\sqrt{5}$. 
Although, this inequality is composed of probabilities, Eq.~(\ref{main_pro}) can be transformed to the inequality with 
the corresponding numbers by using the Azuma's inequality~\cite{Azuma1967}. 
If we apply the Azuma's inequality to the sum of the bit- and phase-error probabilities and the sum of $\tr\hat{\sigma}\hat{P}_1$, 
$\tr\hat{\sigma}\hat{P}_3$ and $\tr\hat{P}_{a\ge2}\hat{\sigma}$ over $N_{\det}$ number of detected events, we have
\begin{align}
e_{\ph}\le \lambda e_{\bit}+
\lambda\sqrt{\frac{N^{N_{\det}}_{a=1}}{N_{\det}}\frac{N^{N_{\det}}_{a=3}}{N_{\det}}}+\frac{N^{N_{\det}}_{a\ge2}}{N_{\det}}
\end{align}
in the asymptotically large $N_{\U{det}}$. 
Here, $N^{N_{\det}}_a$ (with $1\le a\le 3$) denotes the number of detected events when Alice obtains the outcome $a$ 
by measuring system $A$. As $\{N^{N_{\det}}_a\}^3_{a=1}$ are not experimentally available, 
we need to take their upper-bounds using $M^{N_{\em}}_a$ (with $1\le a\le 3$) that represents the number of emitted blocks 
when Alice obtains the outcome $a$ by measuring system $A$. In so doing, we obtain
\begin{align}
\label{eq_maineh}
e_{\ph}\le \lambda e_{\bit}+
\frac{\lambda}{Q}\sqrt{\frac{M^{N_{\em}}_{a=1}}{N_{\em}}\frac{M^{N_{\em}}_{a=3}}{N_{\em}}}+\frac{1}{Q}\frac{M^{N_{\em}}_{a\ge2}}{N_{\em}}.
\end{align}
Once we obtain the following upper bounds
\begin{align}
\Pr\{\wt(\bm{z})=3\}&\le s^U_3,
\label{Pua3}\\
\Pr\{\wt(\bm{z})=1\}&\le s^U_1,
\label{Pua1}\\
\Pr\{\wt(\bm{z})\ge2\}&\le s^U_{\ge2},	
\label{Pua2}
\end{align}
which are determined in the next subsection, the Chernoff bound gives
\begin{align}
\frac{M^{N_{\em}}_{a=3}}{N_{\em}}\le s^U_{3}+\chi,\label{eq1}\\
\frac{M^{N_{\em}}_{a=1}}{N_{\em}}\le s^U_{1}+\chi,\label{eq2}\\
\frac{M^{N_{\em}}_{a\ge2}}{N_{\em}}\le s^U_{\ge2}+\chi.\label{eq3}
\end{align}
If the number $N_{\em}$ of emitted blocks gets larger for
any fixed $\chi>0$, the probability of violating each inequality decreases exponentially. 
Under the asymptotic limit of large $N_{\em}$, we neglect $\chi$ in the following discussions. 
By substituting Eqs.~(\ref{eq1})-(\ref{eq3}) to Eq.~(\ref{eq_maineh}), we finally obtain
\begin{align}
e_{\ph}\le \lambda e_{\bit}+\frac{s^U_{\ge2}}{Q}+
\frac{\lambda\sqrt{s^U_1s^U_{3}}}{Q},
\label{afterchernoff}
\end{align}
which ends the proof of Theorem~\ref{mainth}.

\subsection{Derivations of $s^U_3, s^U_1$ and $s^U_{\ge2}$}
Here, we prove that $s^U_3, s^U_1$ and $s^U_{\ge2}$ can be expressed as Eqs.~(\ref{eq_sU3}), (\ref{eq_sU1}) 
and (\ref{eq_sU2}), respectively. 

\subsubsection{Derivation of $s^U_3$}
Let $n_{\U{block}}:=\sum^3_{j=1}n_j$  denote the number of photons in a single-block of pulses with 
$n_j$ being the number of photons contained in system $S_j$. Then, $\Pr\{\wt(\bm{z})=3\}$ is calculated as
\begin{align}
&\Pr\{\wt(\bm{z})=3\}=\sum^2_{n_\U{block}=0}\Pr\{n_\U{block},\wt(\bm{z})=3\}\notag\\
+&\sum^{\infty}_{n_\U{block}=3}\Pr\{n_\U{block},\wt(\bm{z})=3\}\notag\\
\le&\sum^2_{n_\U{block}=0}\Pr\{n_\U{block},\wt(\bm{z})=3\}+q_3,
\label{maincU3}
\end{align}
where we use Bayes' theorem in the the equality and use Eq.~(\ref{eq_pron}) in the inequality.  

Below, we calculate $\Pr\{n_\U{block},\wt(\bm{z})=3\}$ for $n_{\U{block}}\in\{0,1,2\}$. 
We first calculate $\Pr\{n_\U{block}=0,\wt(\bm{z})=3\}$: 
\begin{align}
&\Pr\{n_\U{block}=0,\wt(\bm{z})=3\}\notag\\
=&\prod^3_{j=1}\U{Pr}\{n_j=0, z_j=1\}\notag\\
=&\prod^3_{j=1}\Pr\{n_j=0|z_j=1\}\Pr\{z_j=1\}.
\label{eq_bayes}
\end{align}
From Eq.~(\ref{coherentLstates}), if $z_j=1$, the $j^{\U{th}}$ state of systems $S_j$ and $R_j$ can be written as 
\begin{align}
\ket{\Phi_-}_{S_jR_j}=(\ket{\psi_0}_{S_jR_j}-\ket{\psi_1}_{S_jR_j})/\mathcal{N}
\label{phi-}
\end{align}
with 
\begin{align}
|\mathcal{N}|^2=2(1-\U{Re}\expect{\psi_1|\psi_0}).
\label{normali}
\end{align}
For state $\ket{\Phi}_{ASR}$ in Eq.~(\ref{coherentLstates}), the probability of obtaining $z_j=1$ is expressed as
\begin{align}
\Pr\{z_j=1\}=|\mathcal{N}|^2/4.
\label{pzj=1}
\end{align}
By expanding the orthonormal basis of system $S_j$ with the photon number states, 
$\ket{\psi_0}_{S_jR_j}$ and $\ket{\psi_1}_{S_jR_j}$ can be respectively written as
\begin{align}
&\ket{\psi_0}_{S_jR_j}\nonumber\\
=&\sqrt{P^{\vac}_{j,0}}\ket{\vac}_{S_j}\ket{u_0}_{R_j}+\sqrt{P^{1}_{j,0}}\ket{1}_{S_j}\ket{u_1}_{R_j}+\cdots,
\label{psi0}
\end{align}
\begin{align}
&\ket{\psi_1}_{S_jR_j}\nonumber\\
=&\sqrt{P^{\vac}_{j,1}}\ket{\vac}_{S_j}\ket{v_0}_{R_j}+\sqrt{P^{1}_{j,1}}\ket{1}_{S_j}\ket{v_1}_{R_j}+\cdots
\label{psi1}
\end{align}
with 
\begin{align}
P^{\vac}_{j,b^A_j}:=\tr\hat{\rho}^{b^A_j}_{S_j}\ket{\vac}\bra{\vac}_{S_j}
\end{align}
and
\begin{align}
P^{1}_{j,b^A_j}:=\tr\hat{\rho}^{b^A_j}_{S_j}\ket{1}\bra{1}_{S_j}.
\end{align}
Here, $\ket{u_0}, \ket{u_1}, \ket{v_0}$ and $\ket{v_1}$ are normalized vectors of system $R_j$. 
Since a purification $\ket{\psi_{b_j^A}}_{S_jR_j}$ 
of $\hat{\rho}^{b_j^A}_{S_j}$ has a freedom of choosing a unitary on system $R_j$, we take $\ket{v_0}$ as $\ket{v_0}=\ket{u_0}$. 
Using Eqs.~(\ref{phi-}), (\ref{psi0}) and (\ref{psi1}) derives 
\begin{align}
\Pr\{n_j=0|z_j=1\}=\frac{(\sqrt{P^{\vac}_{j,0}}-\sqrt{P^{\vac}_{j,1}})^2}{|\mathcal{N}|^2}.
\label{nj0pz1}
\end{align}
By substituting Eqs.~(\ref{pzj=1}) and (\ref{nj0pz1}) to Eq.~(\ref{eq_bayes}), we obtain 
\begin{align}
\Pr\{n_\U{block}=0,\wt(\bm{z})=3\}
=\prod^3_{j=1}\left(\frac{\sqrt{P^{\vac}_{j,0}}-\sqrt{P^{\vac}_{j,1}}}{2}\right)^2.
\end{align}
Through characterization 
of the light source, as we have the bounds on the vacuum emission probabilities, the following inequality holds for any $j\in\{1,2,3\}$. 
\begin{align}
&\Pr\{n_j=0,z_j=1\}=
\left(\frac{\sqrt{P^{\vac}_{j,0}}-\sqrt{P^{\vac}_{j,1}}}{2}\right)^2\notag\\
\le&\frac{\max\left\{(\sqrt{p^U_0}-\sqrt{p^L_1})^2,(\sqrt{p^L_0}-\sqrt{p^U_1})^2\right\}}{4}
=:t
\label{useful}
\end{align}
Note that if the vacuum emission probabilities are equal as assumed in~\cite{npjdps2019}, $P^{\vac}_{j,0}=P^{\vac}_{j,1}$ holds and hence $t=0$. 
Therefore, Eq.~(\ref{eq_bayes}) leads to
\begin{align}
\Pr\{n_\U{block}=0,\wt(\bm{z})=3\}\le t^3.
\label{nb0wt3}
\end{align}	
Next, we calculate $\Pr\{n_\U{block}=1, \wt(\bm{z})=3\}$. From Eq.~(\ref{useful}), we obtain
\begin{align}
&\Pr\{n_\U{block}=1,\wt(\bm{z})=3\}\notag\\
\le&\Pr\{n_1=0,z_1=1\}\Pr\{n_2=0,z_2=1\}\notag\\
+&\Pr\{n_1=0,z_1=1\}\Pr\{n_3=0,z_3=1\}\notag\\
+&\Pr\{n_2=0,z_2=1\}\Pr\{n_3=0,z_3=1\}\notag\\
\le&3t^2.
\label{nbiwt3}
\end{align}
Finally, we calculate $\Pr\{n_\U{block}=2,\wt(\bm{z})=3\}$. 
\begin{align}
&\Pr\{n_\U{block}=2,\wt(\bm{z})=3\}\notag\\
\le&\Pr\{n_1=0,z_1=1\}\Pr\{n_2=0,z_2=1\}\notag\\
+&\Pr\{n_1=0,z_1=1\}\Pr\{n_3=0,z_3=1\}\notag\\
+&\Pr\{n_2=0,z_2=1\}\Pr\{n_3=0,z_3=1\}\notag\\
+&\sum^3_{j=1}\Pr\{n_j=0,z_j=1\}\notag\\
\le&3t^2+3t.
\label{nb2wt3}
\end{align}
Substituting Eqs.~(\ref{nb0wt3}), (\ref{nbiwt3}) and (\ref{nb2wt3}) to Eq.~(\ref{maincU3}) gives
\begin{align}
\Pr\{\wt(\bm{z})=3\}\le q_3+t^3+6t^2+3t=:s_3^U.
\label{uz3}
\end{align}

\subsubsection{Derivation of $s^U_1$}
Here, we prove that $s_1^U$ is expressed as Eq.~(\ref{eq_sU1}). 
\begin{align}
&\Pr\{\wt(\bm{z})=1\}=\Pr\{n_\U{block}=0,\wt(\bm{z})=1\}\notag\\
+&\sum^{\infty}_{n_\U{block}=1}\Pr\{n_\U{block},\wt(\bm{z})=1\}\notag\\
\le&\Pr\{n_\U{block}=0,\wt(\bm{z})=1\}+q_1,
\label{mainwtz1}
\end{align}
where we use Bayes' theorem in the the equality and use Eq.~(\ref{eq_pron}) in the inequality.  

Below, we calculate $\Pr\{n_\U{block}=0,\wt(\bm{z})=1\}$. 
From Eq.~(\ref{useful}), we obtain
\begin{align}
\Pr\{n_\U{block}=0,\wt(\bm{z})=1\}&\le\sum^3_{j=1}\Pr\{n_j=0,z_j=1\}\notag\\
&\le3t.
\label{nb0wt1}
\end{align}
Substituting Eq.~(\ref{nb0wt1}) to Eq.~(\ref{mainwtz1}) gives
\begin{align}
\Pr\{\wt(\bm{z})=1\}\le q_1+3t=:s^U_1.
\label{uz1}
\end{align}

\subsubsection{Derivation of $s^U_{\ge2}$}
Here, we prove that $s_{\ge2}^U$ is expressed as Eq.~(\ref{eq_sU2}). 
\begin{align}
&\Pr\{\wt(\bm{z})\ge2\}=\sum^{1}_{n_\U{block}=0}\Pr\{n_\U{block},\wt(\bm{z})\ge2\}\notag\\
+&\Pr\{n_\U{block}\ge2,\wt(\bm{z})\ge2\}\notag\\
\le&\sum^{1}_{n_\U{block}=0}\Pr\{n_\U{block},\wt(\bm{z})\ge2\}+q_2,
\label{mainge2}
\end{align}
where we use Bayes' theorem in the the equality and use Eq.~(\ref{eq_pron}) in the inequality. 
Below, we calculate $\Pr\{n_\U{block},\wt(\bm{z})\ge2\}$ for $n_\U{block}\in\{0,1\}$. 
From Eq~(\ref{nb0wt3}), we have 
\begin{align}
&\Pr\{n_\U{block}=0,\wt(\bm{z})\ge2\}\notag\\
\le&\Pr\{n_\U{block}=0,\wt(\bm{z})=2\}+t^3.
\end{align}
By using Eq.~(\ref{useful}), it is straightforward to show that $\Pr\{n_\U{block}=0,\wt(\bm{z})=2\}$ is upper bounded as 
\begin{align}
&\Pr\{n_\U{block}=0,\wt(\bm{z})=2\}\notag\\
\le&\Pr\{n_1=0,z_1=1\}\Pr\{n_2=0,z_2=1\}\notag\\
+&\Pr\{n_1=0,z_1=1\}\Pr\{n_3=0,z_3=1\}\notag\\
+&\Pr\{n_2=0,z_2=1\}\Pr\{n_3=0,z_3=1\}\notag\\
\le&3t^2.
\end{align}
Hence, 
\begin{align}
\Pr\{n_\U{block}=0,\wt(\bm{z})\ge2\}\le3t^2+t^3.
\label{subn0wtge2}
\end{align}
Finally, we calculate $\Pr\{n_\U{block}=1,\wt(\bm{z})\ge2\}$. 
From Eq.~(\ref{nbiwt3}), we obtain
\begin{align}
&\Pr\{n_\U{block}=1,\wt(\bm{z})\ge2\}\notag\\
\le&\Pr\{n_\U{block}=1,\wt(\bm{z})=2\}+3t^2.
\end{align}
By using Eq.~(\ref{useful}), 
$\Pr\{n_\U{block}=1,\wt(\bm{z})=2\}$ is upper bounded as
\begin{align}
\Pr\{n_\U{block}=1,\wt(\bm{z})=2\}
\le3(2t+t^2).
\end{align}
Hence, 
\begin{align}
\Pr\{n_\U{block}=1,\wt(\bm{z})\ge2\}\le6t+6t^2.
\label{subn1wtge2}
\end{align}
Substituting Eqs.~(\ref{subn0wtge2}) and (\ref{subn1wtge2}) to Eq.~(\ref{mainge2}) gives the upper bound on 
$\Pr\{\wt(\bm{z})\ge2\}$ as
\begin{align}
\Pr\{\wt(\bm{z})\ge2\}\le q_2+t^3+9t^2+6t=:s^U_{\ge2}.
\label{uz2}
\end{align}

\end{document}